\documentclass[aps,prl,twocolumn,showpacs,amsmath,amssymb,superscriptaddress]{revtex4-1}

\usepackage[utf8]{inputenc}

\usepackage{graphicx}
\usepackage[colorlinks=true,citecolor=blue]{hyperref}
\usepackage{units}

\usepackage{verbatim} 

\newcommand{\ket}[1]{|#1\rangle}
\renewcommand{\vec}[1]{\mathbf{#1}}
\newcommand{\kBT}{k_\text{B}T}

\begin{document}

\title{Mesoscopic Stoner instability in metallic nanoparticles revealed by shot noise}

\author{Bj\"orn Sothmann}
\affiliation{D\'epartement de Physique Th\'eorique, Universit\'e de Gen\`eve, CH-1211 Gen\`eve 4, Switzerland}
\author{J\"urgen K\"onig}
\affiliation{Theoretische Physik, Universit\"at Duisburg-Essen and CeNIDE, 47048 Duisburg, Germany}
\author{Yuval Gefen}
\affiliation{Dept. of Condensed Matter Physics, Weizmann Institute of Science, Rehovot 76100, Israel}

\date{\today}

\begin{abstract}
We study sequential tunneling through a metallic nanoparticle close to the Stoner instability coupled to parallely magnetized electrodes.
Increasing the bias voltage successively opens transport channels associated with excitations of the nanoparticle's total spin.
For the current this leads just to a steplike increase.
The Fano factor, in contrast, shows oscillations between large super-Poissonian and sub-Poissonian values as a function of bias voltage. 
We explain the enhanced Fano factor in terms of generalized random-telegraph noise and propose the shot noise as a convenient tool to probe the mesoscopic Stoner instability. 
\end{abstract}


\maketitle

\paragraph{Introduction.--}
The onset of ferromagnetic long-range order in bulk metals was proposed by Stoner~\cite{stoner_collective_1936} as a consequence of the competition between kinetic and exchange energy. 
The Stoner instability leads to a state with a macroscopically large total spin $S$.
This may be different in mesoscopic systems such as metallic nanoparticles or large quantum dots, which
can be modeled by a universal Hamiltonian characterizing their discrete single-particle level spacing, Coulomb interaction, and exchange interaction of the electron spins.
The latter gives rise to the so called mesoscopic Stoner instability.
The nanoparticle becomes partially polarized and the value of the total spin $S$ sensitively depends on the ratio of exchange interaction and level spacing~\cite{kurland_mesoscopic_2000,aleiner_quantum_2002}.
Recent theoretical studies of the mesoscopic Stoner regime addressed the statistics of linear conductance peaks~\cite{alhassid_effects_2003}, the interplay with Coulomb charging~\cite{kiselev_interplay_2006,burmistrov_spin_2010} and the Kondo effect~\cite{murthy_interplay_2005,rotter_interacting_2008}, the influence of spin-orbit interaction~\cite{gorokhov_combined_2004,tuereci_spin-orbit_2006}, the competition with superconducting pairing~\cite{falci_interplay_2003,schmidt_mesoscopic_2008}, its sensitivity to the degree of chaos~\cite{ullmo_quantum-dot_2003} and its implications on the thermopower~\cite{billings_signatures_2010}. 

An experimental confirmation of the mesoscopic Stoner instability is difficult for at least two reasons.
First, magnetic moments of individual nanoparticles or quantum dots are hard to measure.
Second, the exchange coupling of a given nanoparticle is not tunable, which would be desirable for studying the successive transitions from zero to full polarization. 
So far, only the behavior of the Coulomb peaks as a function of magnetic field served as a tool to probe the spin properties of quantum dots~\cite{duncan_coulomb-blockade_2000,folk_ground_2001,lindemann_stability_2002,potok_spin_2003}.

Transitions between different values $S$ of the total spin can also be achieved by tunnel coupling the nanoparticle to source and drain electrodes with a finite bias voltage $V$.
Increasing the bias voltage enables transitions between different values $S$ of the total spin. 
In this Letter, we suggest to identify the mesoscopic Stoner instability by using ferromagnetic electrodes with parallel magnetization directions and measuring the shot noise.
We find that for an even number of accessible values for the total spin $S$, the Fano factor is strongly enhanced while for an odd number it stays sub-Poissonian. 
This results in oscillations of the Fano factor as a function of bias voltage that are robust against asymmetries in the tunnel couplings.
We propose these oscillations as a striking evidence of the mesoscopic Stoner instability.

\paragraph{Model and technique.--}

The nanoparticle is represented as a multi-level quantum dot described by the universal Hamiltonian~\cite{kurland_mesoscopic_2000,aleiner_quantum_2002},
\begin{equation}\label{eq:Hdot}
	H_\text{dot}=\sum_{\alpha\sigma}\varepsilon_\alpha c_{\alpha\sigma}^\dagger c_{\alpha\sigma}+E_C(N-N_G)^2-J\vec S^2 \, .
\end{equation}
The first term describes the spin-degenerate single-particle levels $\alpha$ of the quantum dot. 
The second term models Coulomb charging. 
Here, $E_C$ is the scale of the charging energy, $N=\sum_{\alpha\sigma}c_{\alpha\sigma}^\dagger c_{\alpha\sigma}$ is the number of electrons on the dot, and $N_G$ is the equilibrium charge of the dot that can be tuned by a gate voltage. 
Finally, the third term describes a Heisenberg-type spin interaction for the total dot spin $\vec S=\sum_{\alpha\sigma\sigma'}c_{\alpha\sigma}^\dagger \vec\sigma_{\sigma\sigma'}c_{\alpha\sigma'}$ with ferromagnetic exchange coupling $J$. 

In the following, we assume a constant single-particle level spacing $\Delta$ and the following hierarchy of energy scales: $E_C, \Delta, J \gg eV \gtrsim \Delta-J \gg \kBT$.
The bias voltage $V$ defines the energy scale for the available excitations.
The inequality $E_C \gg eV$ implies that only two charge states $N_0$ and $N_0+1$ participate in transport. 
Due to the condition $\Delta \gg eV$ we can neglect all states with particle-hole excitations. 
Hence, the dot eigenstates $\ket{N,S,S_z}$ are fully characterized by the total number $N$ of electrons, 
the total spin $S$ and its $z$-component $S_z$. 
The corresponding eigenenergy, 
\begin{equation}
	E_{N,S}=\Delta\left[\left(\frac{N}{2}\right)^2+\frac{N}{2}+S^2\right]+E_C(N-N_G)^2-JS(S+1) \, , 
\end{equation}
is independent of $S_z$.
For a given number of electrons $N$, the ground state spin $S$ is the integer value (for even $N$) or half-integer (for odd $N$) value that is closest to $J/[2(\Delta -J)]$.
This value increases with increasing $J$ and diverges for $J\rightarrow \Delta$.
The latter marks the (macroscopic) Stoner instability.
The excitation energies for changing the total spin $S$ by one while keeping the charge number $N$ fixed form a ladder with constant spacing $2(\Delta-J)$, i.e., the condition $eV \gtrsim \Delta -J$ ensures that more and more spin states are involved with increasing bias voltage.
Finally, the inequality $\Delta-J \gg \kBT$ guarantees that thermal smearing is small enough to resolve individual states with different total spin $S$. 

To achieve transport, the nanoparticle is tunnel coupled to two ferromagnetic electrodes with parallel magnetizations. The Hamiltonian of the total system is given by
$H=H_\text{dot}+\sum_r H_r+H_\text{tun}$.
The ferromagnets are described in terms of noninteracting electrons at chemical potential $\mu_r$ as $H_r=\sum_{\vec k\sigma} \varepsilon_{\vec k\sigma}a_{r\vec k\sigma}^\dagger a_{r\vec k\sigma}$, where $a_{r\vec k\sigma}^\dagger$ creates an electron with momentum $\vec k$ and spin $\sigma$ in lead $r=\text{L,R}$. We assume the leads to have a constant, spin-dependent density of states $\rho_{r\sigma}$ which is related to the polarization of the leads via $p_r=(\rho_{r+}-\rho_{r-})/(\rho_{r+}+\rho_{r-})$.
In the following, we assume $p_\text{L}=p_\text{R} \equiv p$.
The coupling between the dot and the leads is described by $H_\text{tun}=\sum_{r\vec k\alpha\sigma} t_ra_{r\vec k\sigma}^\dagger c_{\alpha\sigma}+\text{H.c.}$ The tunnel matrix elements $t_r$ are related to the tunnel-coupling strength via $\Gamma_{r\sigma}=2\pi|t_r|^2\rho_{r\sigma}$.
Furthermore, we define $\Gamma_r = \sum_\sigma \Gamma_{r\sigma}/2$.

In order to evaluate the current and the current noise we use a real-time diagrammatic approach~\cite{koenig_zero-bias_1996,koenig_resonant_1996} which takes into account the interactions on the dot exactly and performs a perturbative expansion in the dot-lead coupling. 
The idea of this approach is to integrate out the noninteracting lead electrons and to describe the remaining quantum-dot system in terms of its reduced density matrix. 
It is sufficient to consider only diagonal matrix elements $p_{N,S,S_z}$ which describe the probability to find the system in the eigenstate $\ket{N,S,S_z}$
\footnote{}. 
In the stationary state, they are determined by the master equation
\begin{equation}
	0 = \sum_{N,S,S_z}W_{N';N}^{S',S_z';S,S_z}p_{N,S,S_z} \, ,
\end{equation}
where $W_{\chi' \chi}$ are the transition rates from $|\chi\rangle = |N,S,S_z\rangle$ to $|\chi'\rangle = |N',S',S'_z\rangle$ evaluated to first order in the tunnel couplings $\Gamma_r$.
The rates for an electron entering the dot are given by
\begin{widetext}
\begin{align}
\label{eq:rate}
	W_{N+1;N}^{S+\frac{1}{2},S_z\pm\frac{1}{2};S,S_z}&=\sum_r (1\pm p) \Gamma_r f_r ( E_{N+1,S+\frac{1}{2}}-E_{N,S})\left|\left\langle S,S_z;\frac{1}{2},\pm\frac{1}{2}\left|\vphantom{\frac{1}{2}}\right.S+\frac{1}{2},S_z\pm\frac{1}{2}\right\rangle\right|^2 
\\
	W_{N+1;N}^{S-\frac{1}{2},S_z\pm\frac{1}{2};S,S_z}&=\sum_r (1\pm p) \Gamma_r f_r ( E_{N+1,S-\frac{1}{2}}-E_{N,S})\left|\left\langle S-\frac{1}{2},S_z\pm\frac{1}{2};\frac{1}{2},\mp\frac{1}{2}\left|\vphantom{\frac{1}{2}}\right.S,S_z\right\rangle\right|^2 \nonumber
\end{align}
\end{widetext}
where $\langle S,S_z;\frac{1}{2},\pm\frac{1}{2}|S',S_z'\rangle$ denotes the Clebsch-Gordan coefficient for adding the spin of the incoming spin up/down electron to the initial spin $S,S_z$ to get the final spin $S',S_z'$, and $f_r(x)=1/(e^{(x-\mu_r)/\kBT}+1)$ is the Fermi function.
The electrochemical potentials of the leads are chosen symmetrically, 
$\mu_\text{R} = -\mu_\text{L} = eV/2$,
such that electrons preferably travel from the left to the right lead.
The rate $W_{N;N+1}^{S,S_z;S',S_z'}$ for the reverse process is the same as $W_{N+1;N}^{S',S_z';S,S_z}$  but with $f_r$ being  replaced by $1-f_r$. 

From the form of Eq.~(\ref{eq:rate}), that applies for parallel magnetized electrodes, it follows that the probabilities $p_{N,S,S_z}$ are independent of $S_z$, i.e., the spin state of the nanoparticle remains rotationally invariant even under voltage bias. 
A different relative orientation of the leads' magnetizations, e.g., antiparallel, would break this symmetry and allow for accumulation of a finite dipole moment. 

The current through the system is given by $I=\langle\hat I\rangle=\sum_{N,N',S,S',S_z,S_z'}W_{\phantom{I}N';N}^{IS',S_z';S,S_z}p_{N,S,S_z}$, 
where the current rates $W^I$ are obtained from the transition rates by multiplying with the number of electrons transferred between the dot and leads. 
We use standard techniques to calculate the current noise $S=\int dt \langle \hat I(t)\hat I(0)+\hat I(0)\hat I(t)-2\langle \hat I\rangle ^2\rangle$ in Coulomb-blockade systems for weak tunnel coupling~\cite{hershfield_zero-frequency_1993,korotkov_intrinsic_1994,bagrets_full_2003} which may also be formulated within a diagrammatic language~\cite{thielmann_cotunneling_2005}. From this we obtain the Fano factor defined as $F=S/(2eI)$.

\begin{figure}
	\includegraphics[width=\columnwidth]{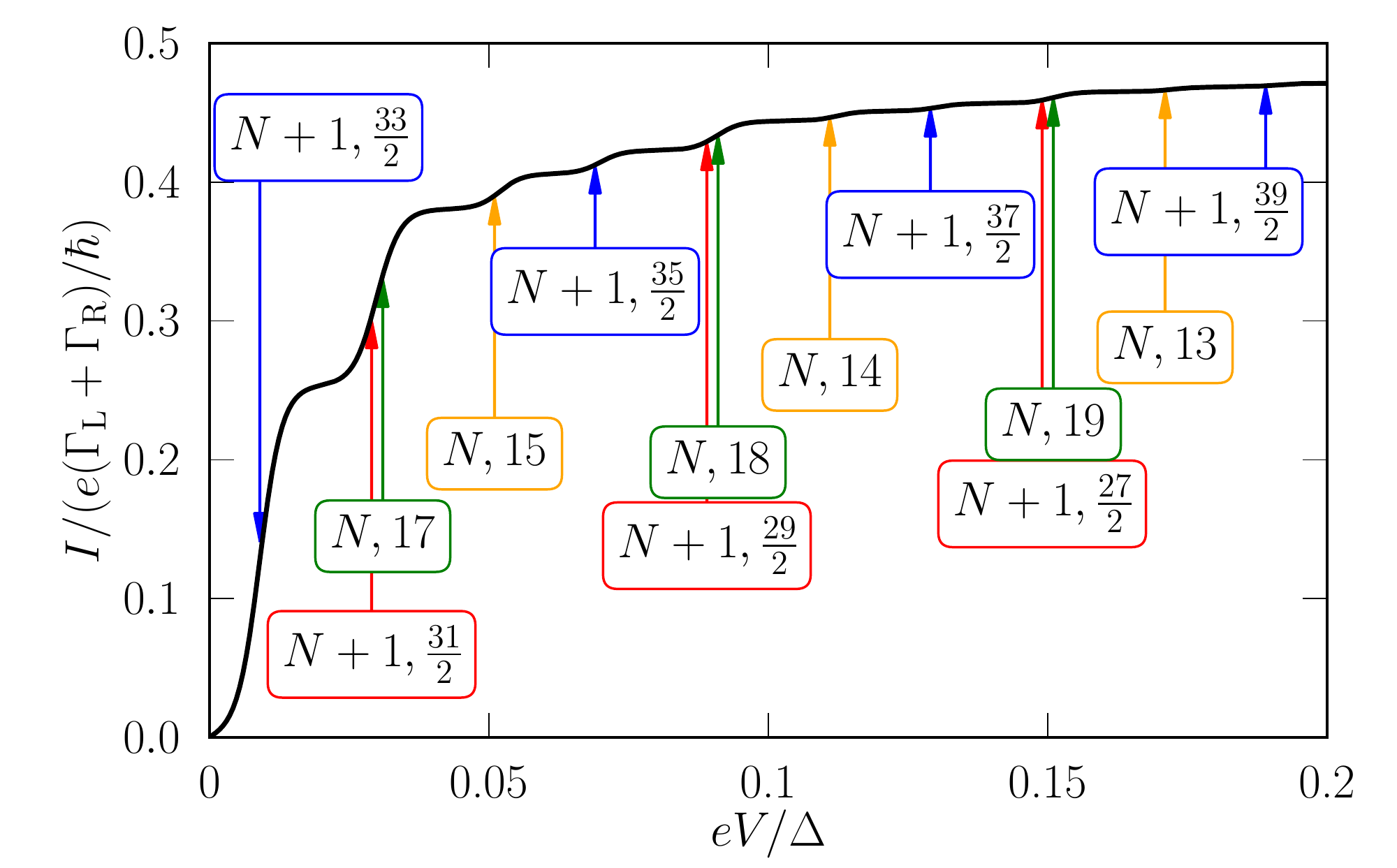}
	\caption{\label{fig:Current}(Color online) Current as a function of bias voltage. Arrows mark onset of transport through excited states. $N_G=n+\frac{1}{2}+\frac{2n+3}{8}\frac{\Delta}{E_C}-0.0001$, $n\in \mathbb{N}$, $E_C=10\Delta$, $J=0.97\Delta$, $\Gamma_\text{L}=\Gamma_\text{R}$, $p=0.3$, $T=0.001\Delta$. Higher steps become increasingly difficult to resolve.}
\end{figure}

\paragraph{Results.--}
\begin{figure}
	\includegraphics[width=\columnwidth]{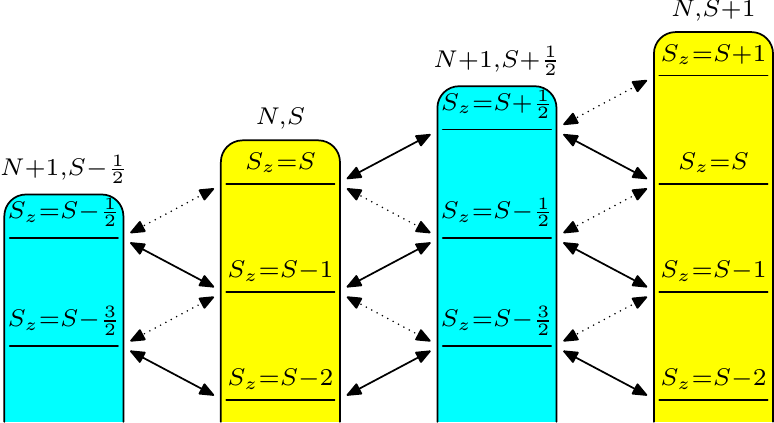}
 	\caption{\label{fig:transport}(Color online) Nanoparticle states and possible transitions. Solid (dotted) lines indicate tunneling of majority (minority) spin electrons.}
\end{figure}
For large charging energy $E_C$, only two charge states, $N$ and $N+1$, are involved in transport.
The number $M$ of possible values for the total spin $S$ increases with increasing bias voltage.
An example for the available states and how they are connected by tunneling is sketched in Fig.~\ref{fig:transport}.
The interesting regime for detecting the mesoscopic Stoner instability is achieved when $N_G$ is tuned (via a gate voltage) in such a way that the lowest excitation energy for a charge transfer, e.g., $E_{N+1,S+1/2}-E_{N,S}$, is smaller than the energy for spin excitations without charge transfer, $E_{N,S\pm 1}-E_{N,S}$, where $N$ and $S$ are the ground state charge and total spin in the absence of bias voltage.
In this case, $M$ increases successively one by one with increasing bias voltage. 
This is accompanied with a step in the current, see Fig.~\ref{fig:Current}, which is smeared by temperature.
Such a series of steps occurs whenever excited states, e.g., of vibrational or magnonic nature~\cite{koch_franck-condon_2005,sothmann_influence_2010}, contribute to transport and is no unique signature of the mesoscopic Stoner regime.

\begin{figure}
	\includegraphics[width=\columnwidth]{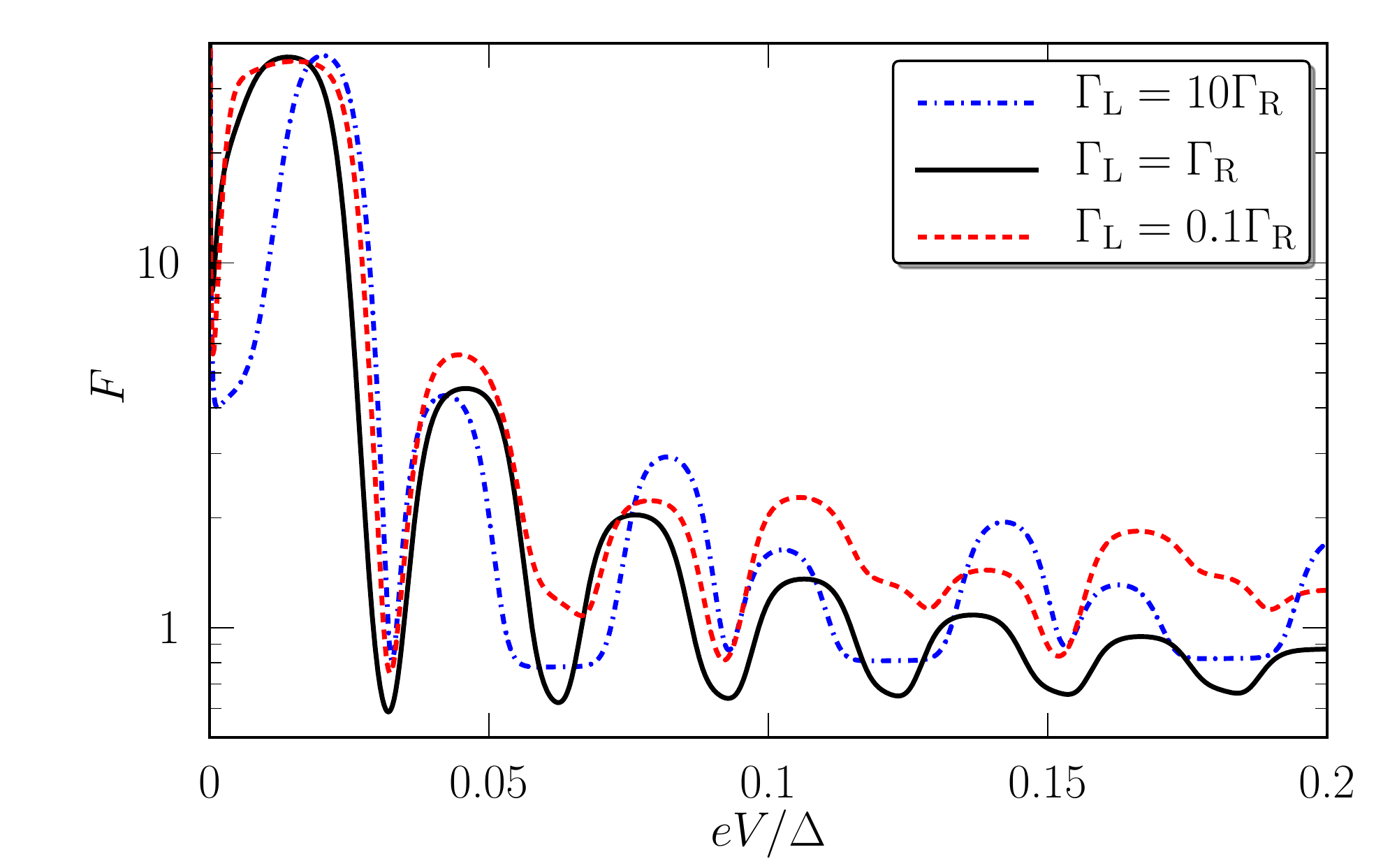}
	\caption{\label{fig:Fano}(Color online) Fano factor as a function of bias voltage for different asymmetries of the tunnel couplings. Other parameters as in Fig.~\ref{fig:Current}.
	}
\end{figure}

Much more interesting behavior is seen in the Fano factor, see Fig.~\ref{fig:Fano}.
The latter oscillates between large super-Poissonian (for even $M$) and sub-Poissonian (for odd $M$) values.
The (zero-temperature) Fano factor of the first plateau ($M=2$), can be written in a compact analytic form.
For the case that only the states $\ket{N,S,S_z}$ and $\ket{N+1,S+\frac{1}{2},S_z}$ contribute to transport, we find 
\begin{multline}
	F(S,p)=\frac{1}{1-p^2}\left\{\frac{\gamma_\text{L}^2+\gamma_\text{R}^2}{(\gamma_\text{L}+\gamma_\text{R})^2}\right.\\\left.+p^2\left[\frac{4S^2}{3}+2S+\left(\frac{4S}{3}+1\right)\frac{\gamma_\text{L}-\gamma_\text{R}}{\gamma_\text{L}+\gamma_\text{R}}\right]\right\},
\end{multline}
where $\gamma_\text{L}=(2S+2)\Gamma_\text{L}$ and $\gamma_\text{R}=(2S+1)\Gamma_\text{R}$.
 
In the limit of unpolarized electrodes, $p=0$, the Fano factor $F=(\gamma_\text{L}^2+\gamma_\text{R}^2)/(\gamma_\text{L}+\gamma_\text{R})^2$ remains sub-Poissonian.
For the limits $S=0$ and $S\to\infty$, we recover the well-known results for transport through a single-level quantum dot, $F=(4\Gamma_\text{L}^2+\Gamma_\text{R}^2)/(2\Gamma_\text{L}+\Gamma_\text{R})^2$~\cite{thielmann_shot_2003} and the result for spinless electrons, $F=(\Gamma_\text{L}^2+\Gamma_\text{R}^2)/(\Gamma_\text{L}+\Gamma_\text{R})^2$~\cite{chen_theoretical_1991}, respectively.

To achieve a super-Poissonian Fano factor, a finite polarization, $p\neq0$, is required.
For $S=0$ we recover the result known for transport through a single-level quantum dot,
$F=[4(1+p^2)\Gamma_\text{L}^2+(1-p^2)\Gamma_\text{R}^2]/[(1-p^2)(2\Gamma_\text{L}+\Gamma_\text{R})^2]$~\cite{braun_frequency-dependent_2006}, which becomes super-Poissonian for $p  > p^*=\sqrt{\Gamma_\text{R}/(2\Gamma_\text{L}+\Gamma_\text{R})}$ and even diverges for $p\to 1$.
The super-Poissonian noise (for $S=0$) is caused by bunching of electrons~\cite{cottet_positive_2004-1,cottet_positive_2004}. 
Majority-spin electrons are transferred through the dot until finally a minority-spin electron enters. 
It blocks further transport as it has a small probability to leave to the drain. 
Therefore, the current gets chopped into bunches of majority-spin electrons.

For probing the mesocopic Stoner instability, we consider the case of finite $p$ and large $S$.
The leading contribution  
\begin{equation}\label{eq:FanoLargeS}
	F=\frac{4}{3} \frac{p^2}{1-p^2}S^2+\mathcal O(S)
\end{equation}
to the Fano factor (of the first plateau) scales quadratically with $S$, diverges for $p \to 1$, and is independent of the tunnel couplings. 
This result can be understood as a generalization of random-telegraph noise.
For this, we consider the different projections $S_z$ for given $N$ and $S$ as the different states among which the system can switch.
As remarked above, the probability distribution $p_{N,S,S_z}$ is $S_z$-independent.
Current is established by transitions that change the state to $\ket{N+1,S+1/2,S_z\pm 1/2}$.
For $p=0$, the two possibilities to increase or decrease the $z$-component of the spin, weighted by the corresponding Clebsch-Gordan coefficients, add up such that for each of the states not only the probability but also the contribution to the current is independent of the initial $S_z$. 
This is different for finite polarization, $p\neq 0$.
Then, transitions involving majority-spin electrons (indicated by solid lines in Fig.~\ref{fig:transport}) are more likely
than those involving minority spins (dotted lines).
As a consequence, the contribution to the current depends on the initial value of $S_z$.
In the considered example, it monotonically increases with $S_z$. 
For a qualitative estimate of the Fano factor, we assume symmetric couplings, $\Gamma_\text{L}=\Gamma_\text{R}$, neglect the Clebsch-Gordan coefficients, and perform a mapping onto an effective two-state system.
The state representing the positive/negative values of $S_z$ carries a current $I_\pm=e(1\pm p/2)\Gamma_\text{L}$.
The average time to change $S_z$ by $1/2$ is $\tau_0= 1/[(1-p^2)\Gamma_\text{L}]$. 
The time $\tau$ to switch from positive to negative $S_z$ by random walk scales with the square of the distance, i.e., $\tau = (2S)^2 \tau_0$.
For frequencies smaller than this switching time, the noise of a random-telegraph signal is given by $(I_+-I_-)^2\tau/2$ \cite{machlup_noise_1954} from which we estimate the Fano factor as $p^2S^2/(1-p^2)$ in qualitative agreement with the result in Eq.~\eqref{eq:FanoLargeS}.

With increasing bias voltage, a third value of the total spin, e.g., $S-1/2$ in addition to $S$ and $S+1/2$, becomes available ($M=3$) and the Fano factor drops to some sub-Poissonian value, see Fig.~\ref{fig:Fano}.
From the alternating way of how majority and minority spins are involved for the transitions it follows that the contribution to the current monotonically increases with $S_z$ for total spin $S+1/2$ but decreases for $S-1/2$, while it is roughly independent of $S_z$ for $S$.
As a consequence, the system can rather quickly change between the high- and low-current states and the random-telegraph noise is reduced.

For larger values of $M$, an $S_z$-dependence of the current is restricted to the smallest ($S_\text{min}$) and largest ($S_\text{max}=S_\text{min}+(M-1)/2$) value of the total spin.
A super-Poissonian Fano factor appears for even $M$, since the current either increases or decreases with $S_z$ for {\em both} $S_\text{min}$ and $S_\text{max}$.
In contrast, the Fano factor remains sub-Poissonian for odd $M$, since the current increases with $S_z$ for $S_\text{min}$ but decreases for $S_\text{max}$, or vice versa.

We finally address the effects of asymmetric tunnel couplings on our results. To this end, we show in Fig.~\ref{fig:Fano} the Fano factor as a function of the applied bias voltage for different asymmetries of the tunnel couplings. Interestingly, the oscillations of the Fano factor persist even for large asymmetries. The asymmetry only has the effect of increasing the Fano factor slightly when many spin states contribute to transport. Thus, the predicted Fano factor oscillations turn out to be very robust which should facilitate their experimental observation.

\paragraph{Conclusions.--}
In conclusion, we discussed spin-polarized transport through a metallic nanoparticle close to the mesoscopic Stoner instability. When only two spin states contribute to transport, we found that random-telegraph noise leads to a super-Poissonian Fano factor that scales with the square of the total spin. 
We, furthermore, found that the Fano factor oscillates as a function of bias voltage showing super (sub-)Poissonian behavior when an even (odd) number of spin states contributes to transport. These oscillations turn out to be robust with respect to asymmetries of the tunnel coupling, thereby offering a good tool to experimentally investigate the mesoscopic Stoner instability.

\paragraph{Acknowledgements.--}
We acknowledge discussions with Moshe Goldstein and financial support from the EU under grant No. 238345 (GEOMDISS), the DFG via SFB 491, the Minerva Foundation, the Israel-Russia MOST grant, and the Israel Science Foundation.


%

\end{document}